\documentclass[11pt,twoside]{book}
\usepackage{konkolyproc2}
\usepackage{longtable}
\usepackage{amsmath,amssymb}
\usepackage{graphicx}
\usepackage{lscape}
\usepackage{index}
\usepackage{natbib}
\usepackage{bigdelim}
\usepackage{multirow}
\makeindex

\begin{document}

\pagestyle{myheadings}
\setcounter{equation}{0}\setcounter{figure}{0}\setcounter{footnote}{0}
\setcounter{section}{0}\setcounter{table}{0}\setcounter{page}{1}
\markboth{Duffau et al.}{Near-field cosmology with RR Lyrae variable stars}
\title{Near-field cosmology with RR Lyrae variable stars: 
A first view of substructure in the southern sky}
\author{S. Duffau$^{1,2}$, A. K. Vivas$^{3}$, C. Navarrete$^{2,1}$, 
M. Catelan$^{2,1}$, G. Hajdu$^{2,1}$, G. Torrealba$^{4}$, C. Cort\'es$^{5,1}$, 
V. Belokurov$^{4}$, S. Koposov$^{4}$, \& A. J. Drake$^{6}$}
\affil{$^1$Millennium Institute of Astrophysics, Santiago, Chile\\
$^2$Instituto de Astrof\'isica, Pontificia Universidad Cat\'olica de Chile, 
 Santiago, Chile\\ 
$^3$Cerro Tololo Inter-American Observatory, La Serena, Chile \\
$^4$Institute of Astronomy, Cambridge CB3 0HA, UK \\
$^5$Departamento de F\'isica, Facultad de Ciencias B\'asicas, Universidad 
Metropolitana de Ciencias de la Educaci\'on,  Santiago, Chile \\
$^6$California Institute of Technology, Pasadena, CA 91225, USA}

\begin{abstract}
We present an update of a spectroscopic follow-up survey at low-resolution 
of a large number of RR Lyrae halo overdensity candidates found in the 
southern sky. The substructure candidates were identified in the RR~Lyrae 
catalog of Torrealba et al. (2015) using Catalina Real-time Transient Survey 
(CRTS) data. Radial velocities and mean metallicities have been estimated 
for target stars in almost half of the original overdensities to assess 
their potential membership to coherent halo features.
\end{abstract}

Surveying the halo for substructure is important to identify the tidal 
remnants of accretion events that contributed to its formation. The northern 
sky has seen already a lot of progress towards this goal (see review by 
Belokurov 2013), but the south remains quite unexplored. The CRTS with the 
work by Torrealba et al. (2015, from now on T15)  has identified 27 
overdensity candidates using RR~Lyrae (RRL) variables. We started a 
low-resolution spectroscopic survey ($R \sim 2000$) using primarily the 
Goodman spectrograph at SOAR and the LDSS3 at Magellan. We obtained spectra 
to measure radial velocities and mean metallicities for the most promising 
targets. Radial velocities are used in addition to positional information to 
search for stream-like features and metallicities for the stream candidate 
stars are measured to properly characterize the nature of the features found 
as real substructures. 

We have covered so far 11 out of the 27 overdensities listed in T15. We have 
obtained more than one spectrum for at least 50\% of our target RRL stars. 
Radial velocities for 123 stars have been measured so far using templates 
from Sesar (2012), and mean metallicities have been estimated using a 
modified version of the $\Delta S$ method developed by Layden (1994) for 
99 of our targets. The typical errors in radial velocity and mean metallicity 
for the sample are $\sim$20 km/s and $\sim$0.2 dex, respectively.

\begin{figure}[!ht]
\includegraphics[width=0.95\textwidth, height=5.7cm]{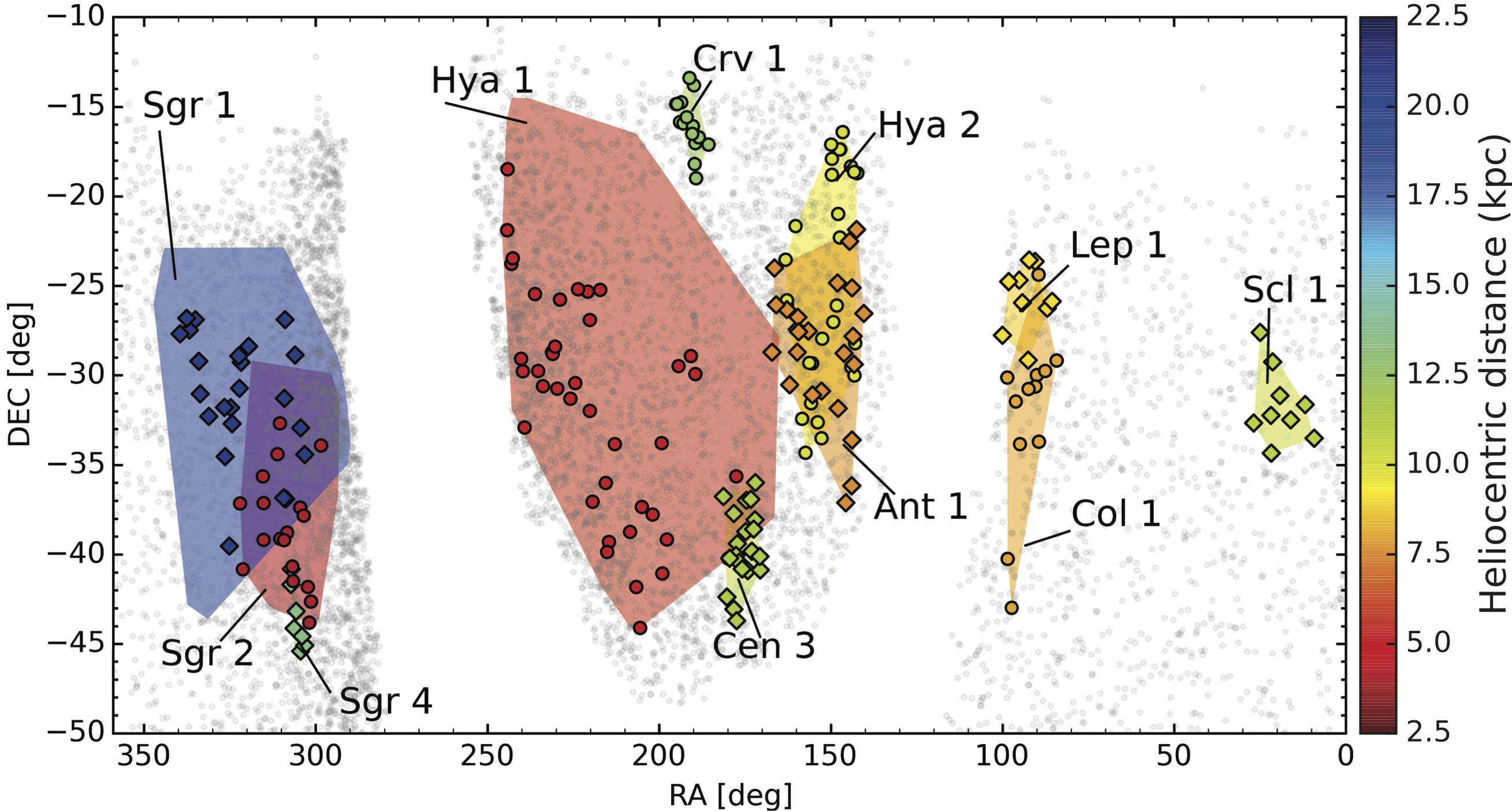}
\vspace*{-2mm}
\caption{Spatial distribution of the overdensities targeted and the 
RR~Lyrae observed within those. See text for more details.} 
\label{fig1} 
\end{figure}

Figure~1 displays the RRLs in the T15 catalog (gray dots) and those observed 
by us (diamonds or ovals) in each of the surveyed overdensities (shaded 
polygons), color-coded according to their mean heliocentric distance. 

Our most interesting preliminary results indicate that we have probably 
found the southern extension of the Virgo Stellar Stream velocity complex 
(Duffau et~al. 2014) in Crv\,1, while a bimodal radial velocity distribution 
in Hya\,2 suggests there is indeed velocity substructure within the 
overdensity. Hya\,1 is interesting as we have proper motion data available 
from the literature to combine with our radial velocities, and because in 
the same direction, albeit at a slightly larger heliocentric distance, a 
large overdensity has been identified by Casetti-Dinescu et~al. (2015) as 
well. Ant\,1 presents an extended radial velocity range with a narrow 
metallicity distribution, while Cen\,3 has a velocity peak like the one 
expected for thick disk stars in the same direction of the sky. 
Combining our data with proper motions, information available from other 
tracers, models of the disruption of the Sagittarius dwarf and the smooth 
halo component of our galaxy will allow us to say more about the nature and 
origin of these overdensities. In particular we will be able to explore the 
relation between the largest overdensity in T15, Sgr\,1, and the tidal 
tails of the Sagittarius dwarf galaxy. 

\vspace*{-2mm}
\section*{Acknowledgments}
\vspace*{-3mm}
Support for this project is provided by CONICYT's PCI program through 
grant DPI\,20140066; by the Ministry for the Economy, Development, and 
Tourism's Iniciativa Cient\'ifica Milenio through grant IC\,120009, 
awarded to the Millennium Institute of Astrophysics; by Proyecto Fondecyt 
Regular \#1141141; and by Proyecto Basal PFB-06/2007. C.N. and G.H. 
gratefully acknowledge support from CONICYT-PCHA/Doctorado Nacional grants 
2015-21151643 and 2014-63140099, respectively.


\vspace*{-2mm}
\begin{thebibliography}{}      
\vspace*{-2mm}
\bibitem[Belokurov (2013)]{B13}
Belokurov, V. 2013, NewAR, 57, 3, 100
\bibitem[Casetti-Dinescu et~al. (2015)]{C15}
Casetti-Dinescu, D.~I., Nusdeo, D.~A., Girard, T.~M., et~al. 2015, \apj, 810, 4
\bibitem[Duffau et~al. (2014)]{D14}
Duffau, S., Vivas, A.~K., Zinn, R., et~al. 2014, A\&A, 566, A118
\bibitem[Layden (1994)]{L94}
Layden, A.~C. 1994, \aj, 108, 1016
\bibitem[Sesar (2012)]{S12}
Sesar, B. 2012, \aj, 144, 114
\bibitem[Torrealba et~al. (2015)]{T15}
Torrealba, G., Catelan, M., Drake, A.~J., et al. 2015, \mnras, 446, 2251
\end{thebibliography}
\end{document}